\documentclass[conference]{IEEEtran}
\IEEEoverridecommandlockouts
\usepackage{cite}
\usepackage{amsmath,amssymb,amsfonts}
\usepackage{algorithmic}
\usepackage{graphicx}
\usepackage{textcomp}
\usepackage{multirow}
\usepackage{xcolor}
\usepackage{colortbl}
\usepackage{lipsum}
\usepackage{url}
\usepackage[bookmarks=false]{hyperref}
\usepackage[most]{tcolorbox}
\usepackage{booktabs}
\usepackage{enumitem}
\usepackage{balance}

\usepackage{xspace}
\newcommand{\al}{\textit{et al.}\xspace}
\newcommand{\ie}{\textit{i.e.,}\xspace}

\definecolor{lightgray}{RGB}{217,217,217}
\definecolor{lightlightgray}{RGB}{235,235,235}

\newcommand{\lightlightgrayrow}{\rowcolor{lightlightgray}}

\newtcolorbox[auto counter]{finding}[1][]{
  enhanced,
  breakable,
  left=2pt,right=2pt,top=1pt,bottom=1pt,colframe=gray!25, arc=1mm, 
  #1
}

\def\BibTeX{{\rm B\kern-.05em{\sc i\kern-.025em b}\kern-.08em
    T\kern-.1667em\lower.7ex\hbox{E}\kern-.125emX}}
\begin{document}

\title{\textit{Log severity levels matter: A multivocal mapping}

\thanks{This research is partially supported by NSERC Discovery Grant and MITACS Acceleration programs.}
}

\author{\IEEEauthorblockN{Eduardo Mendes}
\IEEEauthorblockA{\textit{Université du Québec à Chicoutimi}\\
Saguenay, QC, Canada\\
eduardo.mendes-de-oliveira1@uqac.ca}
\and
\IEEEauthorblockN{Fabio Petrillo}
\IEEEauthorblockA{\textit{Université du Québec à Chicoutimi}\\
Saguenay, QC, Canada\\
fabio@petrillo.com}
}

\maketitle

\begin{abstract}
The choice of log severity level can be challenging and cause problems in producing reliable logging data.
However, there is a lack of specifications and practical guidelines to support this challenge.
In this study, we present a multivocal systematic mapping of log severity levels from literature peer-reviewed, logging libraries, and practitioners' views.
We analyzed 19 severity levels, 27 studies, and 40 logging libraries.
Our results show redundancy and semantic similarity between the levels and a tendency to converge the levels for a total of six levels. Our contributions help leverage the reliability of log entries: (i) mapping the literature about log severity levels, (ii) mapping the severity levels in logging libraries, (iii) a set of synthesized six definitions and four general purposes for severity levels.
We recommend that developers use a standard nomenclature, and for logging library creators, we suggest providing accurate and unambiguous definitions of log severity levels.
\end{abstract}

\begin{IEEEkeywords}
log severity level, logging library, log entry, systematic mapping, multivocal
\end{IEEEkeywords}

\section{Introduction}
Logs are often the primary source of information for system developers and operators to understand and diagnose the behavior of a software system \cite{IST/EL2020/systematic}. According to Lin \al \cite{ICSE/LIN2016/log-clustering}\textit{ “engineers need to examine the  recorded  logs  to  gain  insight  into  the  failure,  identify  the problems,  and  perform  troubleshooting”}. As reported by El-Masri \al \cite{IST/EL2020/systematic}, each log entry is usually composed of time-stamp, severity level, software component, and log message. Severity levels indicate the degree of severity of the log message \cite{SPE/KIM2020/automatic}. For example, a less severe level is used to indicate that the system behaves as expected, while a more severe level is used to indicate that a problem has occurred \cite{ICSE/CHEN2017/characterizing-antipatterns}.

The choice of severity level impacts the amount of log data that a software system produces
\cite{ICSE/LIN2016/log-clustering}\cite{ICSE/CHEN2017/characterizing-antipatterns}\cite{ESE/CHOWDHURY2018/exploratory}\cite{ESE/ZENG2019/studying}. 
For example, if a system is set to \textit{Warn} level, only statements marked with \textit{Warn} and higher levels (e.g., \textit{Error}, \textit{Fatal}) will be output \cite{ICSE/CHEN2017/characterizing-antipatterns}.

In this sense, when a developer choose severity levels inappropriately, the system can produce more log entries than it should, or the opposite, less log entries \cite{ESE/HASSANI2018/studying}. In both scenarios, the wrong choice of severity level can cause problems in the software system performance \cite{ICSE/CHEN2017/characterizing-antipatterns}  \cite{ESE/LI2017/LogLevelChoose} \cite{ICSE/YUAN2012/characterizingLoggingPractices},
 in the maintenance \cite{ESE/LI2017/LogLevelChoose} \cite{ASE/HE2018/characterizingNaturalLanguageDescriptions}, as well affect log-based monitoring and diagnostics \cite{ESE/HASSANI2018/studying} \cite{ESE/LI2017/LogLevelChoose} \cite{ASWEC/RONG2018/logging}.

Developers spend significant time adjusting log severity levels \cite{ESE/KABINNA2018/examining}. After an initial choice, developers may modify the severity level re-evaluating how critical an event is \cite{ICSE/YUAN2012/characterizingLoggingPractices} \cite{SPSP/ZHAO2017/log20}. They can re-evaluate if a statement, initially classified as \textit{Info}, would actually be of \textit{Error} level, or if it would not be an intermediate level between the two levels, \ie a \textit{Warn} \cite{SPSP/ZHAO2017/log20}. 
Among the factors that make choosing the severity level a challenge are: 
(i) lack of knowledge of how logs will be used \cite{COMACM/OLINER2012/advances}; 
(ii) lack of understanding how critical an event is \cite{ESE/ZENG2019/studying}; 
(iii) the ambiguity of certain events that seem to be related to multiple levels of severity \cite{ICSE/LIN2016/log-clustering}\cite{SPSP/ZHAO2017/log20}.

In addition, there is \textit{a lack of specifications and practical guidelines} for performing logging tasks in projects and industry \cite{ASE/HE2018/characterizingNaturalLanguageDescriptions} \cite{ASWEC/RONG2018/logging} \cite{ICSME/ANU2019/verbosityloglevels}. The consequence is that, in software development projects, \textit{“personal experience and preferences play an important role in logging practices”}\cite{ASWEC/RONG2018/logging}.

Considering the lack of guidelines and specifications for logging practices, we found studies in the literature that focus on “where to log” \cite{SPSP/ZHAO2017/log20}\cite{ICSE/FU2014/developers}\cite{TSE/LI2020/qualitative} and “what to log” \cite{ESE/LI2017/LogLevelChoose}. However, we came across a gap in studies that specifically analyze the log severity levels. Thus, we address the following research question: \textit{What are log severity levels?}

To answer this question, we studied the state of the art and practice of log severity levels, surveying their nomenclatures, definitions and descriptions, using three different sources: (1) \textit{peer-reviewed literature}, (2) \textit{logging libraries}, and (3) \textit{practitioners' point of view}.

Our results provide a panorama of log severity levels and show a convergence between academia and industry definitions. We observed that when putting the set of nomenclatures and definitions raised in perspective, we can see a convergence toward four  purposes: \textit{Debugging}, \textit{Informational}, \textit{Warning}, and \textit{Failure}. Furthermore, we proposed definitions for these purposes and the six severity levels that characterize the logging's state of the practice. Our study meets the needs of guidelines and specifications reported in the literature, supporting developers and system operators in generating reliable log data entries. Our study also supports the logging library creators that can use our results to adopt severity levels accordingly.

The main contributions of this study are:
\begin{itemize}
    \item a mapping of the literature on log severity levels;
    \item a mapping of severity levels in the logging libraries;
    \item a set of synthesized definitions for six log severity levels, and four  general purposes for severity levels.
\end{itemize}

This paper is organized as follows. The following Section presents the multivocal mapping of log severity levels covering the peer-reviewed literature, logging libraries and practitioners’ point of view. Section \ref{SECTION/combined-results} presents the log severity levels synthesis. Section \ref{SECTION/discussion} presents a discussion of the main findings, our recommendations,  and the threats to validity. Section \ref{SECTION/conclusion} closes the study presenting our conclusions and future work.

\section{Systematic Mapping of Log Severity Levels}
\label{SECTION/mapping-severity-levels}

We conducted our research on log severity levels using three sources: \textit{peer-reviewed literature} to capture the state of the art; \textit{logging libraries} to capture a vision of library creators; and, to capture \textit{the practitioners' point of view}, a Q/A website. 
All data are available in our reproducibility package at \url{https://github.com/Log-Severity-Level/multivocal-mapping-database}.

\subsection{Methodology - Peer-Reviewed Literature}

\begin{figure}[ht]
\centerline{\includegraphics[width=1\linewidth]{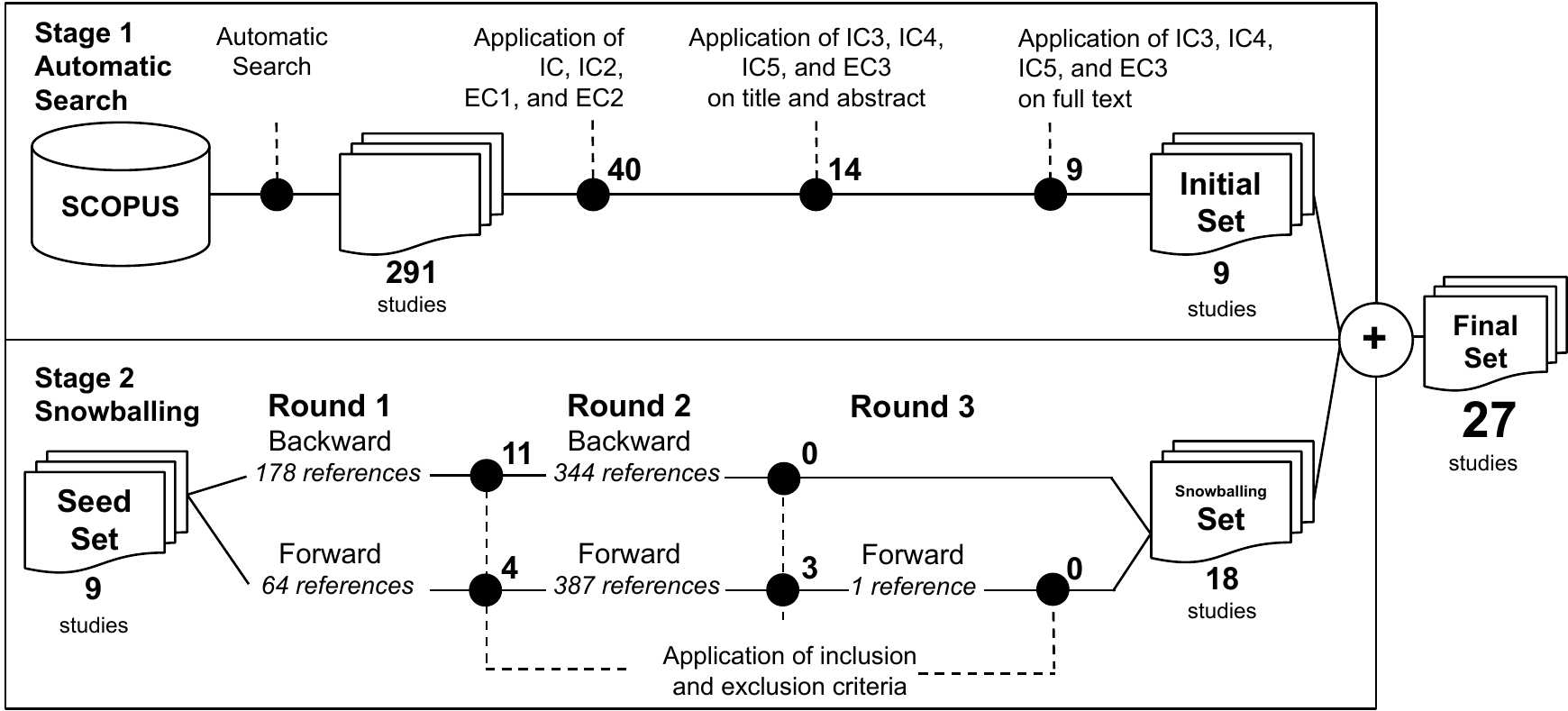}}
\caption{Peer-reviewed selection process.}
\label{IMAGE/peer-reviewed-selection-methodology}
\end{figure}

We performed a two-stage systematic search in order to identify the current literature on log severity levels, as shown in \textbf{Fig. \ref{IMAGE/peer-reviewed-selection-methodology}}. On Stage 1, we adopted automated search as the search strategy. According to Keele \al \cite{KEELE2007/guidelines}, automated search is the most common utilized search strategy to identify relevant studies for a Systematic Mapping. 

In Stage 1, our search query was: \texttt{("log level" OR "log severity" 
 OR "logging level" OR "logging severity"
 OR ("severity level" AND (logging OR log))}. We executed our search query on Scopus\footnote{https://www.scopus.com}, using three metadata fields: title, abstract and keywords and we found \textbf{291 studies}. Then we applied the inclusion (IC) and exclusion (EC) criteria, specifically:

\begin{itemize}
    \item \textbf{IC1:} The study must be a conference paper or article;
    \item \textbf{IC2:} The study must be of the Computer Science area;
    \item \textbf{IC3:} The study must be a primary study;
    \item \textbf{IC4:} The study should address logging practices;
    \item \textbf{IC5:} The study should describe the use of log severity levels or define them;
    \item \textbf{EC1:} The study is not written in English;    
    \item \textbf{EC2:} The study is a duplicate;
    \item \textbf{EC3:} The study does not present a link between logging practices and the use of log severity levels.
\end{itemize}

After applying the IC1, IC2, EC1, and EC2, we obtained \textbf{40 studies}. We read the title and abstract of each of them and, after filtering by IC3, IC4, IC5, and EC3, we kept \textbf{14 studies}. We read all 14 papers, filtering by IC3, IC4, IC5, and EC3; we obtained the seed data set with \textbf{9 studies}.

In Stage 2, we used our seed data set to perform three rounds of snowballing, backward and forward, detailed in \textbf{Fig. \ref{IMAGE/peer-reviewed-selection-methodology}}. 

The final data set included \textbf{27 studies} (\textbf{Table \ref{TABLE/included-studies}}). 

\bgroup
\setlength{\tabcolsep}{0.20em}
\begin{table*}[ht]
\scriptsize
\centering

  \caption{List of included peer-reviewed studies}
  \label{TABLE/included-studies}
  \begin{tabular}{c c l c}
    \toprule
    \# & Reference & Title & Year\\
    \hline
        \lightlightgrayrow {[}P01{]}  & \cite{COMACM/OLINER2012/advances}      & Advances and challenges in log analysis                     & 2012 \\

        {[}P02{]}  & \cite{ICSE/YUAN2012/characterizingLoggingPractices}       & Characterizing logging practices in open-source software                     & 2012 \\

        \lightlightgrayrow {[}P03{]}  & \cite{WCCCT/GOMATHY2014/developing}       & Developing an error logging framework for ruby on rails application using AOP                     & 2014 \\

        {[}P04{]}  & \cite{ICSE/FU2014/developers}       & Where do developers log? An empirical study on logging practices in industry                     & 2014 \\

        \lightlightgrayrow {[}P05{]}  & \cite{ESE/SHANG2015/studying}       & Studying the relationship between logging characteristics and the code quality of platform software                     & 2015 \\

        {[}P06{]}  & \cite{ICSE/LIN2016/log-clustering}       & Log clustering based problem identification for online service systems                     & 2016 \\

        \lightlightgrayrow {[}P07{]}  & \cite{ESE/LI2017/LogLevelChoose}          & Which log level should developers choose for a new logging statement?                             & 2017 \\

        {[}P08{]}  & \cite{ICSE/CHEN2017/characterizing-antipatterns}             & Characterizing and Detecting Anti-Patterns in the Logging Code   & 2017 \\

        \lightlightgrayrow {[}P09{]}  & \cite{ESE/CHEN2017/characterizing-logging-practices}             & Characterizing logging practices in Java-based open source software projects – a replication study in Apache Software Foundation   & 2017 \\

        [P10]  & \cite{SPSP/ZHAO2017/log20}             & Log20: Fully Automated Optimal Placement of Log Printing Statements under Specified Overhead Threshold   & 2017 \\

        \lightlightgrayrow {[}P11{]}  & \cite{ESE/LI2017/towards}             & Towards just-in-time suggestions for log changes   & 2017 \\

        [P12]  & \cite{ASWEC/RONG2018/logging}             & How is logging practice implemented in open source software projects? A preliminary exploration   & 2018 \\

        \lightlightgrayrow {[}P13{]}  & \cite{ASE/HE2018/characterizingNaturalLanguageDescriptions}   & Characterizing the natural language descriptions in software logging statements                  & 2018 \\

        [P14]  & \cite{ESE/HASSANI2018/studying}   & Studying and detecting log-related issues                  & 2018 \\

        \lightlightgrayrow {[}P15{]}  & \cite{ESE/CHOWDHURY2018/exploratory}   & An exploratory study on assessing the energy impact of logging on android applications                  & 2018 \\

        [P16]  & \cite{ESE/KABINNA2018/examining}   & Examining the stability of logging statements                  & 2018 \\

        \lightlightgrayrow {[}P17{]}  & \cite{CLOUD/YUAN2019/approach}   & An approach to cloud execution failure diagnosis based on exception logs in Openstack                  & 2019 \\

        [P18]  & \cite{ICSME/ANU2019/verbosityloglevels}   & An Approach to Recommendation of Verbosity Log Levels Based on Logging Intention                  & 2019 \\

        \lightlightgrayrow {[}P19{]}  & \cite{ESE/ZENG2019/studying}              & Studying the characteristics of logging practices in mobile apps: a case study on F-Droid         & 2019 \\

        [P20] & \cite{ICSE/LI2019/dlfinder}              & DLFinder: Characterizing and Detecting Duplicate Logging Code Smells                                                & 2019 \\

        \lightlightgrayrow {[}P21{]} & \cite{ESE/CHEN2019/extracting}              & Extracting and studying the Logging-Code-Issue-Introducing changes in Java-based large-scale open source software systems                                                & 2019 \\

        [P22] & \cite{TSE/LI2020/qualitative}              & A Qualitative Study of the Benefits and Costs of Logging from Developers Perspectives                                                & 2020 \\

        \lightlightgrayrow {[}P23{]} & \cite{SPE/KIM2020/automatic}              & Automatic recommendation to appropriate log levels                                                & 2020 \\

        [P24] & \cite{ICTSS/BHARKAD2020/optimizing}       & Optimizing Root Cause Analysis Time Using Smart Logging Framework for Unix and GNU/Linux Based Operating System   & 2020 \\

        \lightlightgrayrow {[}P25{]} & \cite{ICDC/OBRKEBSKI2019/log}       & Log Based Analysis of Software Application Operation   & 2020 \\
        
        [P26] & \cite{ACMSAC/GHOLAMIAN2020/logging}       & Logging statements' prediction based on source code clones   & 2020 \\

        \lightlightgrayrow {[}P27{]} & \cite{ACMCASE/LI2020/where-shall}       & Where Shall We Log? Studying and Suggesting Logging Locations in Code Blocks   & 2020 \\

    \bottomrule
  \end{tabular}
\end{table*}
\egroup

\subsection{Results - Peer-Reviewed Literature}

\textbf{Distribution of included studies.} The highest number of publications that include log severity levels were published in the last four years: five studies in 2017, five studies in 2018, five studies in 2019, and six studies in 2020. These 21 studies represent 78\% of our included studies and make us observe that interest in the subject has been increasing in recent years.

Most of the included studies deal with severity levels in general, presenting them to illustrate and clarify logging processes as a whole. Other studies address severity levels as one aspect of their research on logging practices [P02][P09][P12][P19], or in the study of log statements [P13][P16][P26]. Some studies deal with specific problems related to severity levels, where to log [P04][P10], which severity level to choose [P07], and automatic recommendation of severity levels [P23][P27].

\begin{finding}[label=FINDING/trend-research]
\small
\textit{\textbf{Finding \#\ref{FINDING/trend-research}:} Research on log severity levels has grown in recent years.}
\end{finding}

\begin{figure}[htbp]
\centerline{\includegraphics[width=0.92\linewidth]{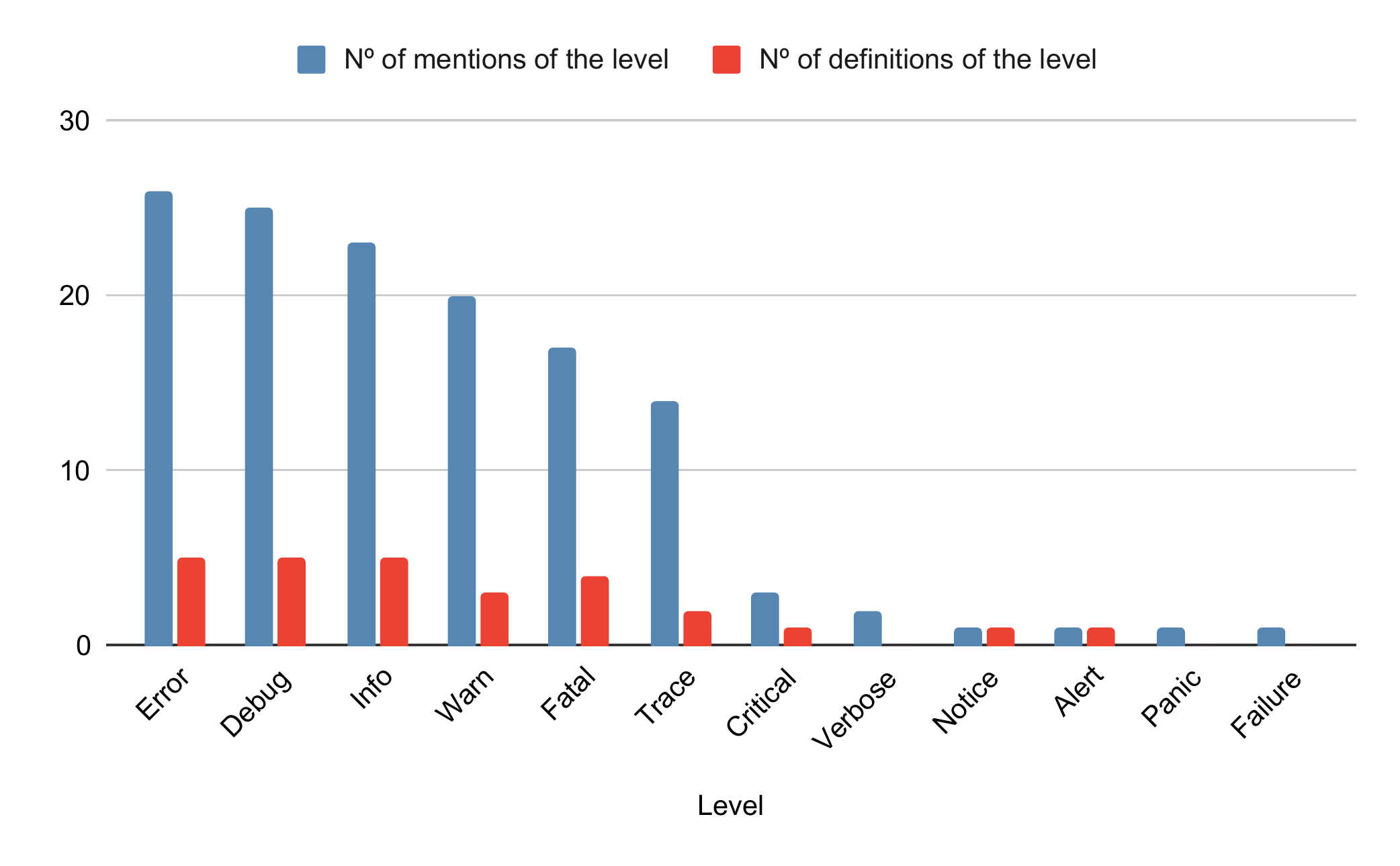}}
\caption{Number of mentions vs number of definitions on literature}
\label{IMAGE/cited-vs-definitions}
\end{figure}

\textbf{Mentions of log severity levels.} All 27 studies mention log severity levels (e.g., \textit{Debug}, \textit{Info}, \textit{Warn},...); 23 studies (85\%) mention at least three  severity levels. In contrast, only eight studies (30\%) have definitions or descriptions for severity levels. 
As shown in \textbf{Fig. \ref{IMAGE/cited-vs-definitions}}, it is possible to distinguish two groups of severity levels: the most mentioned and the least mentioned. The first one formed by the \textit{Error} (26), \textit{Debug} (25), \textit{Info} (23), \textit{Warn} (20), \textit{Fatal} (17), and \textit{Trace} (14) levels, makes up 93\% of the mentions, and the latter formed by the \textit{Notice}, \textit{Critical}, \textit{Alert}, \textit{Verbose}, \textit{Panic}, and \textit{Failure} levels, making up the remaining 7\%.
The most representative group in the number of mentions is also the one that comes with the most definitions.

\begin{finding}[label=FINDING/main-levels-research]
\small
\textit{\textbf{Finding \#\ref{FINDING/main-levels-research}:} Error, Debug, Info, Warn, Fatal, and Trace are the levels that stand out in the log severity level research.}
\end{finding}

\textbf{Categorization.}
Some studies propose categorizations related to log statements. He \al \cite{ASE/HE2018/characterizingNaturalLanguageDescriptions} group the \textit{logging descriptions}\footnote{\textit{“(...) the textual part of a log statement, excluding variables”} \cite{ASE/HE2018/characterizingNaturalLanguageDescriptions}} into three  main categories: (i) \textit{“description for program operation”}, (ii) \textit{“description for error condition”}, and (iii) \textit{“description for high-level code semantics”}. The descriptions of the first category appear related to the \textit{Info} severity level, describing three  types of operations: complete operation, current operation, and next operation. The second category describes the occurrence of an error/exception; severity levels related to this category are \textit{Error} and \textit{Info}. In the third category, the logging descriptions essentially describe the code, e.g., variables, functions, and branches, such as if-else blocks; all examples of this category use \textit{Debug} level.

Yuan \al \cite{ICSE/YUAN2012/characterizingLoggingPractices} comment on two classes of levels: \textit{“error-level (e.g., error, fatal) (...) and non-error (also non-fatal levels), such as info and debug”}. In the same way, Shang \al \cite{ESE/SHANG2015/studying} comment on other two classes considering average logging level\footnote{The “\textit{average logging level}” is a metric calculated from transforming each log severity level into quantitative measures
\cite{ESE/SHANG2015/studying}.}:

\begin{quote}
    \textit{“Intuitively, high-level logs are for system operators and lower-level logs are for development purposes. (...) The higher-level logs are used typically by administrators and lower-level logs are used by developers and testers”.} \cite{ESE/SHANG2015/studying}
\end{quote}

\begin{finding}[label=FINDING:studies-and-categories]
\small
\textit{\textbf{Finding \#\ref{FINDING:studies-and-categories}:} Studies seek to categorize the severity levels or elements of log sentences.}
\end{finding}

\textbf{Definitions}.
\textbf{Table \ref{TABLE/severity-levels-definition-papers}} presents the definitions and descriptions found for the log severity levels. Four studies have definitions for a defined set of levels (four or more levels) [P02][P12][P14][P24].
The other four studies present descriptions for only one or two levels to contextualize the idea of log severity level [P20][P23][P25][P27].

\begin{finding}[label=FINDING:low-number-definitions]
\small
\textit{\textbf{Finding \#\ref{FINDING:low-number-definitions}:} Only 15\% of selected studies have a set of four  or more definitions for log severity levels.}
\end{finding}

Next, we comment on the levels with more than one definition in Table \ref{TABLE/severity-levels-definition-papers}.

\paragraph{\textbf{Debug}} As described by its name, four  of the definitions associate the \textit{Debug} level with debugging tasks [P02][P12][P14][P24]. Its target phase of the software process is development, consequently consumed mainly by developers [P23][P24]. \textit{Debug} level appears related to expressions like \textit{“verbose,” “fine-grained information,” “details of events,” “useful for developers.”} Kim \al \cite{SPE/KIM2020/automatic}  describe it as \textit{“broadly used to designate the state of the variable.”}

\paragraph{\textbf{Trace}} The \textit{Trace} level is described as \textit{“more finer grained”} than the \textit{Debug} level [P12][P14].

\paragraph{\textbf{Info}} \textit{Info} level messages are described as \textit{“important but normal events”} [P02] [P14][P24], used to highlight and describe the application's progress [P12][P25] \textit{“at coarse-grained level,”} whose circumstances do not require action to take [P24].

\paragraph{\textbf{Warn}} Unlike the \textit{Info} level, the \textit{Warn} level definitions describe it as a severity level that requires action to be taken [P24] because it  designates potentially harmful situations [P12] capable of causing system problems [P27]. 

\paragraph{\textbf{Error}} The definitions for the \textit{Error} level do not say much beyond their goal of logging errors or failed operations [P02][P12][P14][P24]. However to [P12], \textit{Error} level \textit{“designates error events that might still allow the application to continue running.”}

\paragraph{\textbf{Fatal}} The expressions used in the \textit{Fatal} level definitions are \textit{“aborting”} processes or applications {[}P02{]}{[}P12{]}{[}P14{]}, \textit{“very severe errors”} {[}P12{]}, and \textit{“critical problems”} {[}P23{]}.

\begin{table*}[ht]
\centering
\setlength{\tabcolsep}{0.20em}
\scriptsize
\caption{Severity level definitions Peer-reviewed literature set}
\begin{tabular}{rcl}
\toprule
Level & Paper & Quote  \\
\hline  
\lightlightgrayrow Trace   & {[}P12{]} & “Trace / Finest: This level designates finer-grained informational events than the 'Debug'." \\
        \lightlightgrayrow & {[}P14{]} &  “and trace (tracing steps of the execution, most fine-grained information)" \\

Debug   & {[}P02{]} & “debug (i.e., verbose logging only for debugging)"  \\
        & {[}P12{]} & “Debug / Fine / Finer: This level designates fine-grained informational events that are most useful to debug an application."  \\
        & {[}P14{]} & “debug (verbose logging only for debugging)"  \\
        & {[}P23{]} & “Moreover, the log level debug is broadly used to designate the state of the variable during the development phase with the corresponding message." \\
        & {[}P24{]} & “[...] Debug: Messages at debug level contains more details of events, debug level log messages are more useful for developers and for debugging an application." \\

\lightlightgrayrow Info    & {[}P02{]} & “info (i.e., record important but normal events)"  \\
        \lightlightgrayrow & {[}P12{]} & “Info / Config: This level designates informational messages that highlight the progress of the application coarse-grained level."  \\
        \lightlightgrayrow & {[}P14{]} & “info (record important but normal events)"  \\
        \lightlightgrayrow & {[}P24{]} & “[...]  Information: Normal operation messages are at this level, no action is required to take." \\
        \lightlightgrayrow & {[}P25{]} & “Info level entries describe application operation, e.g. details of creating services." \\

Notice  & {[}P24{]} & “Level 5 Notice: Unusual event is mentioned, but not, an error is shown." \\

\lightlightgrayrow Warn    & {[}P12{]} & “Warn / Warning: This level designates potentially harmful situations."  \\
        \lightlightgrayrow & {[}P24{]} & “[...] Warning: Warning messages indicate that an error may occur if action is not taken."  \\
        \lightlightgrayrow & {[}P27{]} & “The logging statement is at the warn level, which is the level for recording information that may potentially cause system oddities"\\

Error   & {[}P02{]} & “error (i.e., record error events)  \\
        & {[}P12{]} & “Error / Severe: This level designates error events that might still allow the application to continue running."  \\
        & {[}P14{]} & “error (record error events)"  \\
        & {[}P20{]} & “The logging statement is at the error level, which is the level for recording failed operations."\\
        & {[}P24{]} & “[...] Error: Occurred error information is shown in this kind of log messages."\\

\lightlightgrayrow Critical & {[}P24{]} & “[...] Critical: Critical level messages, is written in the log file when a critical situation occurs in the normal execution of the system."  \\
Alert   & {[}P24{]} & “[...] Alert: Alert level messages indicate that respective one should be corrected immediately."  \\

\lightlightgrayrow Fatal   & {[}P02{]} & “fatal (i.e., abort a process after logging)" \\
        \lightlightgrayrow & {[}P12{]} & “Fatal: This level designates very severe error events that will presumably lead the application to abort."  \\
        \lightlightgrayrow & {[}P14{]} & “fatal (abort a process after logging)" \\
        \lightlightgrayrow & {[}P23{]} & “For example, the log level fatal is used to indicate that a critical problem has occurred around the position of the log statement, where the developer \\
        \lightlightgrayrow & & tries to leave an appropriate log message as a clue to treat it later. " \\
\bottomrule
\multicolumn{3}{l}{$^{\mathrm{*}}$The levels are distributed from top to bottom, from the least severe to the most severe.}\\
\end{tabular}
\label{TABLE/severity-levels-definition-papers}
\end{table*}

\subsection{Methodology - Logging Libraries}

We use the PYLP index\footnote{https://pypl.github.io/PYPL.html}, a ranking of programming languages, as a starting point for library selection, selecting languages with a “share value” greater than 1.0\%. We got \textbf{16 languages}, so we took these languages and queried  Google Search: \texttt{logging library}, concatenating the name of each language and used the first result page for each query. We found \textbf{160 hits} (blogs, forums, code repositories), and from them, we mapped \textbf{60 libraries}. We inspect code repositories (when available), documentation and library guidelines to apply our inclusion and exclusion criteria:

\begin{itemize}
    \item \textbf{IC1:} The library/language has a set of log severity levels;
    \item \textbf{EC1:} The library does not create log statements with log severity levels;
    \item \textbf{EC2:} The library is on Github and has less than 1000 stars.
\end{itemize}

After applying the above criteria, we obtained \textbf{37 libraries}. We manually added Java Util Logging, PHP logging, and Syslog-ng to the set. Our final data set included \textbf{40 libraries}\footnote{Among the libraries, three appear in two versions (Log4J [L13], versions 1 and 2; Loguru versions C++[L27] and Python [L34]; PHP Logging, versions Linux and Windows [L36])} \textbf{(Table \ref{TABLE/severity-levels-on-logging-libraries})} and \textbf{63 documents} among source code, documentation and library guidelines.

\subsection{Results - Logging Libraries}

\textbf{Table \ref{TABLE/severity-levels-on-logging-libraries}} presents the logging libraries selected for the study and the \textbf{19 different severity levels} found in them. The levels are distributed from left to right, from the least severe to the most severe. The table does not show the pseudo-levels (e.g., \textit{All, Off, Notset, Log4Net\_Debug} [L17][L40]) and groups the variant nomenclatures for the same level (\textit{Info/Informational}, and \textit{Warn/Warning}).

\textbf{Programming languages.} The included libraries cover 14 of the 16 selected programming languages. The most significant number of libraries are from C/C++ with seven libraries (20\%), Java with six libraries (15\%), JavaScript with five  libraries (13\%), and C\# and Golang, both with four  libraries (10\%).

\textbf{Distribution of levels by library.} Of the selected libraries, 91\% have between five and eight severity levels, 39.5\% have six levels, 23.3\% have five levels, 14\% have eight levels, and 14\% have seven levels. The libraries with the lowest number of levels, Google Glog [L01] and Golang Glog [L02], have the four  same levels: \textit{Info, Warn, Error}, and \textit{Fatal}. The libraries with more levels are Log4C [L39] and Log4Net [L40], with nine and 15 severity levels, respectively.

\begin{finding}[label=FINDING/min-max-number-libraries]
\small
\textit{\textbf{
Finding \#\ref{FINDING/min-max-number-libraries}:} Logging libraries has a median of six severity levels ($\sigma \approx 1.8$). The lowest number of log severity levels in libraries is four, and the highest number is fifteen.
}
\end{finding}

\textbf{Occurrence of levels.} When aggregating the data from Table \ref{TABLE/severity-levels-on-logging-libraries}, we observe that six levels are present in more than 50\% of the libraries, four levels of which are present in more than 90\%: \textit{Info} (100\%), \textit{Warn} (98\%), \textit{Error} (98\%), \textit{Debug} (93\%), \textit{Trace} (55\%), and \textit{Fatal} (52\%).

\begin{finding}[label=FINDING/main-levels-libraries]
\small
\textit{\textbf{Finding \#\ref{FINDING/main-levels-libraries}:} Six levels are present in over 50\% of libraries, among them four in over 90\%: \textit{Info} (100\%), \textit{Warn} (98\%), \textit{Error} (98\%), \textit{Debug} (93\%), \textit{Trace} (55\%), and \textit{Fatal} (52\%).}
\end{finding}

Four libraries have one severity level that is unique to them: \textit{Fault} in [L03], \textit{Config} in [L29], \textit{Basic} in [L32], and \textit{Success} in [L34]. Another six levels are only present in up to 10\% of libraries: \textit{Verbose} (10\%), \textit{Emergency} (10\%), \textit{Finer} (10\%), \textit{Finest} (7\%), \textit{Fine} (5\%), and \textit{Severe} (5\%).

\begin{finding}[label=FINDING/low-levels-libraries]
\small
\textit{\textbf{Finding \#\ref{FINDING/low-levels-libraries}:} Basic, Config, Emergency, Fault, Fine, Finest, Finer, Severe, Success, and Verbose have low occurrence in libraries compared to other severity levels, less than or equal to 10\%.}
\end{finding}

\begin{figure}[b]
\centerline{\includegraphics[width=1\linewidth]{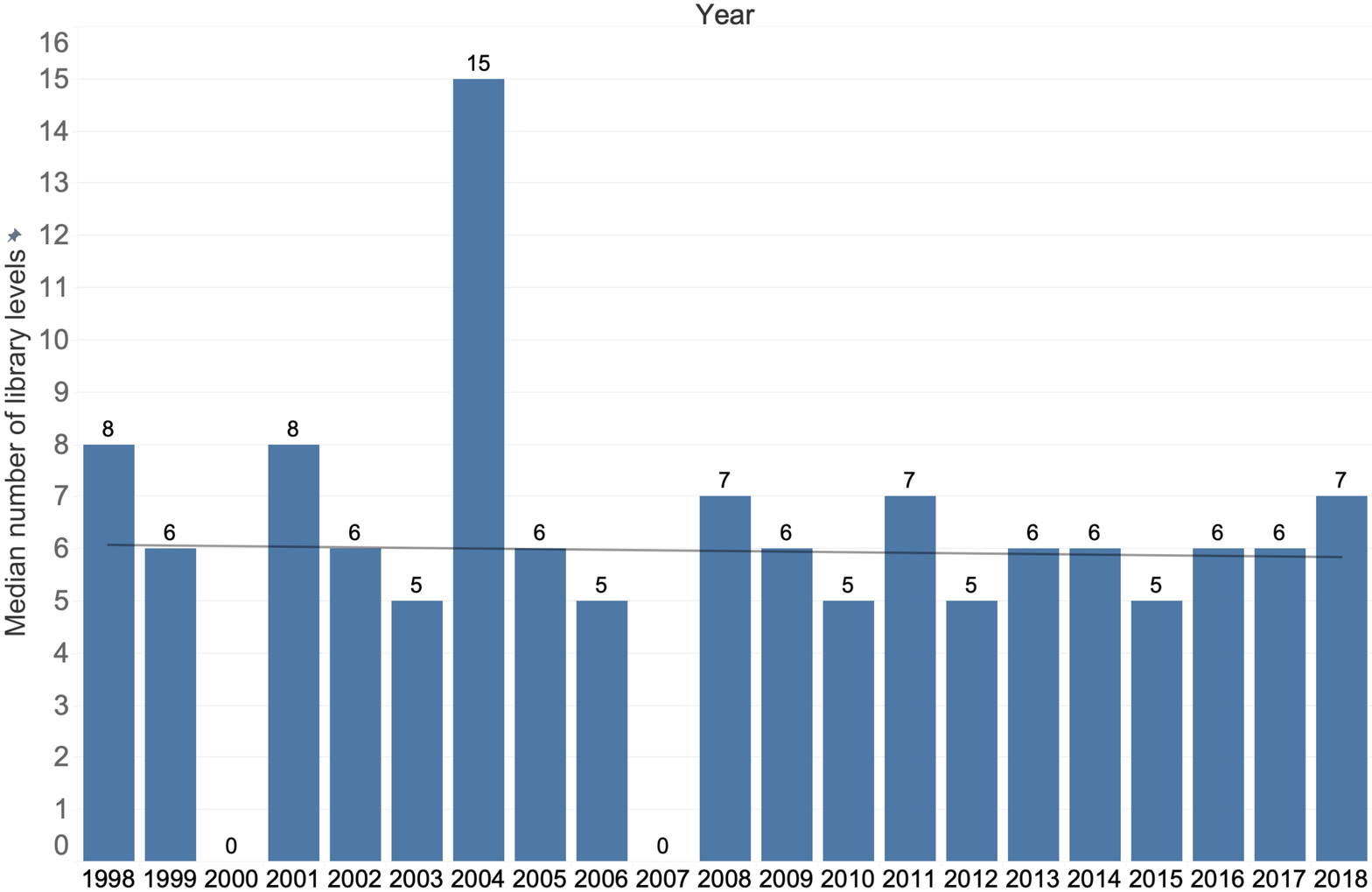}}
\caption{Year of release and number of levels: The data collected suggest that there is no trend towards an increase in the severity levels (The value 0 indicates the absence of new libraries in the related year).}
\label{IMAGE/libraries-years-vs-totals}
\end{figure}

\textbf{Number of severity levels over time.} \textbf{Fig. \ref{IMAGE/libraries-years-vs-totals}} presents the medians of the number of levels of libraries by their release years. There is no linear variation over time in the number of severity levels that the logging libraries have provided. There is even a slight variation in quantity. The only point outside the curve is the year 2004, which features the 15 levels of the Log4Net library. Despite this high number, its documentation informs that it \textit{“categorizes logging into levels: DEBUG, INFO, WARN, ERROR and FATAL”} [L40]. 

\begin{finding}[label=FINDING/trend-number-libraries]
\small
\textit{\textbf{Finding \#\ref{FINDING/trend-number-libraries}:} There is no trend towards a decrease or increase in the number of log severity levels in future libraries.}
\end{finding}

\textbf{The most severe level.} The most severe level in 48\% of the libraries is the \textit{Fatal} level, followed by the \textit{Error} (19\%), \textit{Critical} (14\%), \textit{Emergency} (10\%), and \textit{Alert} (5\%) levels. The eight libraries, where the \textit{Error} level is the most severe, have five logging severity levels.

\textbf{Numeric Values.} In our data set, 38 libraries (95\%) use numeric values associated with levels to sort them according to their specific degrees of severity (\textbf{Table \ref{TABLE/severity-levels-on-logging-libraries}}), as indicated by the following quote: 

\begin{quote}
    \textit{“Levels have a numeric value that defines the relative ordering between levels.”} [L40]
\end{quote}

\bgroup
\setlength{\tabcolsep}{0.13em}
\begin{table*}[]
\centering
\scriptsize
\caption{Severity Levels on Logging Libraries}
\label{TABLE/severity-levels-on-logging-libraries}

\begin{tabular}{c l c c c c c c c c c c c c c c c c c c c c c}
\toprule
\# & Library & Launch & Nº & Finest & Verbose & Finer & Trace & Debug & Basic & Fine & Config & Info & Success & Notice  & Warn & Error & Fault & Severe & Critical & Alert & Fatal & Emerg. \\
\hline

{[}L01{]} & Google Glog$^{\mathrm{a}}$ & 2015 & 4  &  &  & &  & &  &  &  &  0 &  &  & 1 & 2 &  & &  &  & 3 &\\
\lightlightgrayrow {[}L02{]} & Golang Glog$^{\mathrm{b}}$ & 2013 & 4 & &  &  &  &  &  &  &  & 0 &  & & 1 & 2 &  & &  &  & 3 & \\
{[}L03{]} & OSLogging$^{\mathrm{c,i}}$ & ? & 5 &  &  &  &  & \checkmark &  & & & \checkmark &  & \checkmark & & \checkmark &  \checkmark &  &  & \\

\lightlightgrayrow {[}L04{]} & Rust Lang$^{\mathrm{d}}$ & 2014 & 5 &  &  &  & 5 & 4 &  &  & & 3 &  & & 2 & 1 &  &  &  &  &  & \\
{[}L05{]} & Logback$^{\mathrm{e}}$ & 2006 & 5 &  &  &  & 5000 & 10000 &  &  & & 20000 &  & & 30000 & 40000 &  &  &  &  &  & \\
\lightlightgrayrow {[}L06{]} & SLF4J$^{\mathrm{e}}$ & 2006 & 5 &  &  &  & 0 & 10 &  &  & & 20 &  & & 30 & 40 &  &  &  &  &  & \\

{[}L07{]} & Ruby Logger$^{\mathrm{f}}$ & 2003 & 5 &  &  &  &  & 0 &  &  & & 1 &  & & 2 & 3 &  &  & &  & 4 &  \\
\lightlightgrayrow{[}L08{]} & LogLevel$^{\mathrm{g}}$ & 2013 & 5 &  &  &  & 0 & 1 &  &  & & 2 &  & & 3 & 4 &  &  &  &  &  & \\
{[}L09{]} & JS-logger$^{\mathrm{g}}$ & 2012 & 5 &  &  &  & 1 & 2 &  &  & & 3 &  & & 5 & 8 &  &  &  &  &  & \\
\lightlightgrayrow {[}L10{]} & CocoaLum.$^{\mathrm{c}}$ & 2010 & 5 &  & 4 &  &  & 3 &  &  & & 2 &  & & 1 & 0 &  &  &  &  &  & \\
{[}L11{]} & Kotlin-logging$^{\mathrm{h}}$ & 2016 & 5 &  &  &  & \checkmark & \checkmark &  &  &  & \checkmark &  & & \checkmark & \checkmark &  &  &  &  &  & \\
\lightlightgrayrow {[}L12{]} & SwiftyBeaver$^{\mathrm{i}}$ & 2015 & 5 &  & 0 &  &  & 1 &  &  & & 2 &  & & 3 & 4 &  &  &  &  &  & \\

\multirow{2}{*}{{[}L13{]}} & \multirow{2}{*}{Log4J$^{\mathrm{e}}$} & 1999 (v1) & \multirow{2}{*}{6} &  &  &  & 5000 & 10000 &  &  & & 20000  &  & & 30000  & 40000  &  &  &  & & 50000  &   \\
 &  & 2014 (v2) &  &  &  &  & 600 & 500 &  &  & & 400 &  & & 300 & 200 &  &  &  & & 100 &   \\

\lightlightgrayrow {[}L14{]} & Commons &  &  &  & &  &  &  &  &  &  &  &  &  &  &  &  &  &  &  &  &  \\
   \lightlightgrayrow & Logging$^{\mathrm{e}}$ & \multirow{-2}{*}{2002} & \multirow{-2}{*}{6} &  &  &  & \multirow{-2}{*}{1} & \multirow{-2}{*}{2} &  &  & & \multirow{-2}{*}{3} &  & & \multirow{-2}{*}{4} & \multirow{-2}{*}{5} &  &  & & & \multirow{-2}{*}{6} & \\
{[}L15{]} & Bunyan$^{\mathrm{g}}$ & 2011 & 6 &  &  &  & 10 & 20 &  &  & & 30 &  & & 40 & 50 &  &  &  & & 60 & \\
\lightlightgrayrow {[}L16{]} & NLog$^{\mathrm{j}}$ & 2006 & 6 &  &  &  & 0 & 1 &  &  & & 2 &  & & 3 & 4 &  &  &  & & 5 & \\
{[}L17{]} & Python & \multirow{2}{*}{2002} & \multirow{2}{*}{6} &  &  &  &  & \multirow{2}{*}{10} &   & & & \multirow{2}{*}{20} &  & & \multirow{2}{*}{30} & \multirow{2}{*}{40} &  & & \multirow{2}{*}{50}  & & \multirow{2}{*}{50} &  \\
& Lang$^{\mathrm{k}}$ & & & & & & & & & & & & & & & & & &\\

\lightlightgrayrow {[}L18{]} & ME Logging$^{\mathrm{l}}$ & 2015 & 6 &  &  &  & 0 & 1 &  &  & & 2 &  & & 3 & 4 &  & & 5 &  &  & \\
{[}L19{]} & Serilog$^{\mathrm{l}}$ & 2013 & 6 &  & 0 &  &  & 1 & & & & 2 &  & & 3 & 4 &  &  &  & & 5 &   \\
\lightlightgrayrow {[}L20{]} & Log4PHP$^{\mathrm{o}}$ & 2009 & 6 &  &  &  & 500 & 10000 &  &  & & 20000 &  & & 30000 & 40000 &  &  & & & 50000 &   \\
{[}L21{]} & PinoJS$^{\mathrm{g}}$ & 2016 & 6 &  &  &  & 10 & 20 &  &  & & 30 &  & & 40 & 50 &  &  &  &  &60 &   \\
\lightlightgrayrow {[}L22{]} & Log.c$^{\mathrm{a}}$ & 2017 & 6 &  &  &  & 0 & 1 &  &  & & 2 &  & & 3 & 4 &  &  & &  & 5 & \\
{[}L23{]} & C-Logger$^{\mathrm{a}}$ & 2016 & 6 &  &  &  & 0 & 1 &  &  & & 2 &  & & 3 & 4 &  &  &  &. & 5 & \\
\lightlightgrayrow {[}L24{]} & Zlog$^{\mathrm{a}}$ & 2011 & 6 &  &  &  &  & 20 &  &  & & 40 & & 60 & 80 & 100 & & &  &  & 120 & \\
{[}L25{]} & Go-logging$^{\mathrm{b}}$ & 2013 & 6 &  &  &  &  & 5 &  &  & & 4 & & 3 & 2 & 1 & & & 0 &  &  & \\
\lightlightgrayrow {[}L26{]} & Log4m$^{\mathrm{m}}$ & ~2012 & 6 &  &  &  & 1 & 2 &  & & & 3 &  & & 4 & 5 & & &  &  & 6 & \\
{[}L27{]} & Loguru$^{\mathrm{a}}$ & 2015 & 6 &  &  &  &  & 6* &  &  &  & 5 &  & & 4 & 3 &  & & & 1 & 2 & \\
\lightlightgrayrow {[}L28{]} & Spdlog$^{\mathrm{a}}$ & 2014 & 6 &  &  &  & 0 & 1 &  &  & & 2 &  & & 3 & 4 & & & 5 &  &  & \\

{[}L29{]} & Java Util  &  &  &  &  & &  &  &  &  &  &  &  &  &  &  &  &  &   &\\
   & Logging$^{\mathrm{e}}$ & \multirow{-2}{*}{2002} & \multirow{-2}{*}{7} & \multirow{-2}{*}{300} &  & \multirow{-2}{*}{400} &  &  & & \multirow{-2}{*}{500} & \multirow{-2}{*}{700} & \multirow{-2}{*}{800} &  & & \multirow{-2}{*}{900} &  &  & \multirow{-2}{*}{1000} &  &  &   &\\

\lightlightgrayrow {[}L30{]} & Logrus$^{\mathrm{b}}$ & 2013 & 7 &  &  &  & 6 & 5 &  &  & & 4 &  & & 3 & 2 &  &  & & & 1 &  \\
{[}L31{]} & Uber-go/zap$^{\mathrm{b}}$ & 2016 & 7 &  &  &  &  & 0 &  &  & & 1 &  & & 2 & 3 &  &  &  &  &  & \\
\lightlightgrayrow {[}L32{]} & Bolterauer$^{\mathrm{n}}$ & 2008 & 7 & 7 &  & 6 &  & & 1 & 5 & & 4 &  & & 3 &  &  &  & & & 2 & \\
{[}L33{]} & Swift-log$^{\mathrm{i}}$ & 2018 & 7 &  &  &  & 0 & 1 &  &  & & 2 & & 3 & 4 & 5 & & & 6 &  &  &   \\
\lightlightgrayrow {[}L34{]} & Loguru$^{\mathrm{k}}$ & 2017 & 7 &  &  &  & 5 & 10 &  &  & & 20 & 25 & & 30 & 40 & & & 50 &  &  &  \\

{[}L35{]} & Syslog-ng$^{\mathrm{a}}$ & 1998 & 8 &  &  &  &  & 7 &  &  & & 6 & & 5 & 4 & 3 & & & 2 & 1 &  & 0 \\
\lightlightgrayrow {[}L36{]} & PHP$^{\mathrm{o}}$ & 2001 &  & & & & & &  &  &  &  &  &  &  &  &  &   &  &  &   &\\
  \lightlightgrayrow  & (on Linux) & & 8 &  &  &  &  & 7 &  &  & & 6 & & 5 & 4 & 3 &  & & 2 & 1 &  & 0  \\
  \lightlightgrayrow  & (on Wind.) & & 8 &  &  &  &  & 6 &  &  & & 6 & & 6 & 5 & 1 & & & 1 & 1 &  & \\

{[}L37{]} & Monolog$^{\mathrm{o}}$ & 2011 & 8 &  &  &  &  & 100 &  & &  & 200 & & 250 & 300 & 400 &  & & 500 & 550 & 600 & \\
\lightlightgrayrow {[}L38{]} & Winston$^{\mathrm{g}}$ & 2011 & 8 &  &  &  &  & 7 &  &  & & 6 & & 5 & 4 & 3 &  & & 2 & 1 &  & 0 \\
{[}L39{]} & Log4C$^{\mathrm{a}}$ & 2002? & 9 &  &  &  & 800 & 700 &  &  & & 600 & & 500 & 400 & 300 &  & & 200 & 100 & 0 &  \\

\lightlightgrayrow {[}L40{]} & Log4Net$^{\mathrm{j}}$ & 2004 & 15 & 10000 & 10000 & 20000 & 20000 & 30000 & & 30000 & & 40000 & & 50000 & 60000 & 70000 & & 80000 & 90000 & 100000 & 110000 & 120000  \\

\hline
 &  &  &  & 3 & 4 & 3 & 23 & 39 & 1 & 3 & 1 & 42 & 1 & 10 & 41 & 41 & 1 & 2 & 13 & 8 & 22 & 4 \\
\bottomrule
    & \multicolumn{19}{l}{$^{\mathrm{a}}$C/C++,
    $^{\mathrm{b}}$Golang,
    $^{\mathrm{c}}$Objective-C,
    $^{\mathrm{d}}$Rust,
    $^{\mathrm{e}}$Java,
    $^{\mathrm{f}}$Ruby,
    $^{\mathrm{g}}$JavaScript,
    $^{\mathrm{h}}$Kotlin,
    $^{\mathrm{i}}$Swift,
    $^{\mathrm{j}}$C\#,
    $^{\mathrm{k}}$Python,
    $^{\mathrm{l}}$.NET,
    $^{\mathrm{m}}$MatLab,
    $^{\mathrm{n}}$VBA,
    $^{\mathrm{o}}$PHP}\\
\end{tabular}
\end{table*}
\egroup

\textbf{Table \ref{TABLE/severity-levels-on-logging-libraries}} shows that it is possible to organize the severity levels of almost all libraries equivalently when considering their numerical values, except for libraries [L17] and [L32]. [L17] presents the only variation in the numerical ordering between \textit{Alert} and \textit{Critical}. [L32] has a \textit{Debug} level numbering different from all other libraries. Furthermore, 60\% of libraries (25) have their levels sorted in ascending order and 31\% (11) in descending order.

\begin{finding}[label=FINDING/consistency-numeric-libraries]
\small
\textit{\textbf{Finding \#\ref{FINDING/consistency-numeric-libraries}:} There is consistency in the numerical ordering of severity levels across logging libraries.}
\end{finding}

\begin{quote}
    \textit{“Two Levels with the same value are deemed to be equivalent.”} [L40]
\end{quote}

It is possible to observe libraries with the same numerical value for different severity levels, they are: (i) the \textit{Critical} and \textit{Fatal} levels in the Python library [L17]; (ii) \textit{Info} and \textit{Notice}, (iii) \textit{Error}, \textit{Critical} and \textit{Alert}, in the PHP library [L36], when running on the Windows operating system; (iv) \textit{Finest} and \textit{Verbose}, (v) \textit{Finer} and \textit{Trace}, (vi) \textit{Debug} and \textit{Fine} in Log4Net [L40].
The fact that different severity levels have the same numerical value indicates redundancy of the log levels, which is well exemplified by the following quote:

\begin{quote}
    \textit{“Why doesn't the org.slf4j.Logger interface have methods for the FATAL level?
    The Marker interface
    (...) renders the FATAL level largely redundant. If a given error requires attention beyond that allocated for ordinary errors, simply mark the logging statement with a specially designated marker which can be named ‘FATAL’ (...) ”} {[}L06{]}
\end{quote}

\begin{finding}[label=FINDING/redundancy-on-libraries]
\small
\textit{\textbf{Finding \#\ref{FINDING/redundancy-on-libraries}:} Three  libraries have redundancy in the numeric values of their log severity levels.}
\end{finding}

The \textit{“Marker interface”}, mentioned in [L06], is an option provided by libraries {[}L06{]} and {[}L13{]} to add more context to a log statement and avoid redundancy, which allows using only the necessary log levels.

\bgroup
\setlength{\tabcolsep}{0.1em}
\begin{table*}
\centering
\scriptsize
\caption{Severity level definitions from libraries}
\begin{tabular}{rcl}
\toprule
Level & Library & Quote  \\
\hline
\lightlightgrayrow Fine*, & {[}L09{]} & “(...) are intended for relatively detailed tracing. The exact meaning of the three levels will vary between subsystems, but in general, FINEST should\\
\lightlightgrayrow Verb. &  & be used for the most voluminous detailed output, FINER for somewhat less detailed output, and FINE for the lowest volume (..) messages.”  \\
\lightlightgrayrow & {[}L12{]} & “These levels designate fine-grained informational events that are most useful to debug an application.” \\
\lightlightgrayrow Verb. & {[}L14{]} & “Verbose is the noisiest level, rarely (if ever) enabled for a production app.”\\

Trace & {[}L04{]} & “Designates very low priority, often extremely verbose, information.”\\
 & {[}L05{]} & “The TRACE level designates informational events of very low importance.”\\
 & {[}L13{]} & “A fine-grained debug message, typically capturing the flow through the application.” \\
 & {[}L14{]} & “(...) more detailed information. Expect these to be written to logs only.” \\
 & {[}L15{]} & “Logging from external libraries used by your app or very detailed application logging.” \\
 & {[}L16{]} & “For trace debugging; begin method X, end method X.” \\
 & {[}L18{]} & “Logs that contain the most detailed messages. (...) may
contain sensitive application data. (...) should never be enabled in a production environment.”\\
 & {[}L40{]} & “The Trace level designates fine-grained informational events that are most useful to debug an application.” \\
 
\lightlightgrayrow Debug & {[}L04{]} & “Designates lower priority information.” \\
 \lightlightgrayrow & {[}L05{]} & “The DEBUG level designates informational events of lower importance.” \\
 \lightlightgrayrow & {[}L13{]} & “A general debugging event.” \\
 \lightlightgrayrow & {[}L14{]} & “Detailed information on the flow through the system. Expect these to be written to logs only.” \\
 \lightlightgrayrow & {[}L15{]} & “Anything else, i. e. too verbose to be included in ‘info’ level.” \\
  \lightlightgrayrow & {[}L16{]} & “For debugging; executed query, user authenticated, session expired.” \\
 \lightlightgrayrow & {[}L18{]} & “(...) used for interactive investigation during development (...) should primarily contain information useful for debugging and have no long-term value.”\\
 \lightlightgrayrow & {[}L19{]} & “Debug is used for internal system events that are not necessarily observable from the outside, but useful when determining how something happened.”\\
 \lightlightgrayrow & {[}L35{]} & “The message is only for debugging purposes.” \\
 \lightlightgrayrow & {[}L36{]} & “debug-level message.” \\
 \lightlightgrayrow & {[}L40{]} & “The Debug level designates fine-grained informational events that are most useful to debug an application.” \\

Info & {[}L04{]} & “Designates useful information.” \\
 & {[}L05{]} & “The INFO level designates informational messages highlighting overall progress of the application.” \\
 & {[}L13{]} & “An event for informational purposes.” \\
 & {[}L14{]} & “Interesting runtime events (startup/shutdown). Expect these to be immediately visible on a console, so be conservative, and keep to a minimum.”\\
 & {[}L15{]} & “Detail on regular operation.” \\
 & {[}L16{]} & “Normal behavior like mail sent, user updated profile etc.” \\
 & {[}L18{]} & “Logs that track the general flow of the application. These logs should have long-term value.”\\
 & {[}L19{]} & “(...) things happening in the system that correspond to its responsibilities and functions. (...) observable actions the system can perform.”\\
 & {[}L29{]} & “INFO is a message level for informational messages. Typically INFO messages will be written to the console or its equivalent.(...)\\
 & {[}L35{]} & “The message is purely informational.”\\
 & {[}L36{]} & “informational message.” \\
 & {[}L40{]} & “The Info level designates informational messages that highlight the progress of the application at coarse-grained level.” \\
 
\lightlightgrayrow Notice & {[}L03{]} & “Captures information that is essential for troubleshooting problems. For example, capture information that might result in a failure.” \\
 \lightlightgrayrow  & {[}L35{]} & “The message describes a normal but important event.” \\
 \lightlightgrayrow & {[}L36{]} & “normal, but significant, condition.” \\
 \lightlightgrayrow & {[}L40{]} & “The Notice level designates informational messages that highlight the progress of the application at the highest level.” \\

Warn & {[}L04{]} & “Designates hazardous situations.” \\
 & {[}L05{]} & “The WARN level designates potentially harmful situations.” \\
 & {[}L13{]} & “An event that might possible lead to an error.” \\
 & {[}L14{]} & “Use of deprecated APIs, poor use of API, ‘almost’ errors, other runtime situations that are undesirable or unexpected, but not necessarily ‘wrong’.\\
 & {[}L15{]} & “A note on something that should probably be looked at by an operator eventually.” \\
 & {[}L16{]} & “Something unexpected; application will continue.” \\
 & {[}L17{]} & “Houston, we have a \%s, ‘bit of a problem’.” \\
 & {[}L18{]} & “Logs that highlight an abnormal or unexpected event in the application flow, but do not otherwise cause the application execution to stop.”\\
 & {[}L29{]} & “WARNING is a message level indicating a potential problem.
In general WARNING messages should describe events that will be of interest \\
 & &  to end users or system managers, or which indicate potential problems.” \\
 & {[}L35{]} & “Warning conditions / The message is warning.” \\
 & {[}L36{]} & “warning conditions.” \\
 & {[}L40{]} & “The Warn level designates potentially harmful situations.” \\
 
\lightlightgrayrow Error & {[}L04{]} & “Designates very serious errors.” \\
 \lightlightgrayrow & {[}L05{]} & “The ERROR level designates error events which may or not be fatal to the application.” \\
 \lightlightgrayrow & {[}L13{]} & “An error in the application, possibly recoverable.” \\
 \lightlightgrayrow & {[}L14{]} & “Other runtime errors or unexpected conditions. Expect these to be immediately visible on a status console.” \\
 \lightlightgrayrow & {[}L15{]} & “Fatal for a particular request, but the service/app continues servicing other requests. An operator should look at this soon(ish).” \\
 \lightlightgrayrow & {[}L16{]} & “Something failed; application may or may not continue.” \\
 \lightlightgrayrow & {[}L18{]} & “(...) highlight when the current flow of execution is stopped due to a failure. These should indicate a failure in the current activity (...)”\\
 \lightlightgrayrow & {[}L35{]} & “The message describes an error.” \\
 \lightlightgrayrow & {[}L36{]} & “error conditions.” \\
 \lightlightgrayrow & {[}L40{]} & “The Error level designates error events that might still allow the application to continue running.” \\
 
Severe & {[}L40{]} & “The Severe level designates very severe error events.” \\

\lightlightgrayrow Critical & {[}L17{]} & “Houston, we have a \%s, ‘major disaster’.” \\
 \lightlightgrayrow & {[}L18{]} & “Logs that describe an unrecoverable application or system crash, or a catastrophic failure that requires immediate attention”\\
 \lightlightgrayrow & {[}L35{]} & “The message states a critical condition.” \\
 \lightlightgrayrow & {[}L36{]} & “critical conditions.” \\
 \lightlightgrayrow & {[}L40{]} & “The Critical level designates very severe error events. Critical condition, critical.” \\
 
Alert & {[}L35{]} & “Action must be taken immediately ”\\
 
 & {[}L36{]} & “action must be taken immediately.” \\
 & {[}L40{]} & “The Alert level designates very severe error events. Take immediate action, alerts.” \\

 \lightlightgrayrow Fatal & {[}L13{]} & “A severe error that will prevent the application from continuing.” \\
 \lightlightgrayrow & {[}L14{]} & “Severe errors that cause premature termination. Expect these to be immediately visible on a status console.” \\
 \lightlightgrayrow & {[}L15{]} & “The service/app is going to stop or become unusable now. An operator should definitely look into this soon.” \\
 \lightlightgrayrow & {[}L16{]} & “Something bad happened; application is going down.” \\
 \lightlightgrayrow & {[}L40{]} & “The Fatal level designates very severe error events that will presumably lead the application to abort.” \\

Emerg. & {[}L35{]} & “The message says the system is unusable.” \\
 & {[}L36{]} & “system is unusable.” \\ 
 & {[}L40{]} & “The Emergency level designates very severe error events. System unusable, emergencies.” \\
\bottomrule
\end{tabular}
\label{TABLE/severity-levels-definition-libraries}
\end{table*}
\egroup

\textbf{Late Trace.} 
The \textit{Trace} level is present in 55\% of selected libraries, however, analyzing the release notes of these libraries, in at least three  of them, the \textit{Trace} level was not present in the first versions. It was added to Log4J [L13] in 2005, SLF4J [L06] in 2007, and JS-Logger [L09] in 2018. According to the SLF4J FAQ page, \textit{Trace} level was used in several projects
\begin{quote}
    \textit{“to disable logging output from certain classes without needing to configure logging for those classes. (...) in many of cases the TRACE level carried the same similar semantics  meaning as DEBUG.” } {[}L06{]}
\end{quote}

\begin{finding}[label=FINDING/trace-redundant-purpose]
\small
\textit{\textbf{Finding \#\ref{FINDING/trace-redundant-purpose}:} There may be semantic similarity in using the Trace and Debug levels.}
\end{finding}

\textbf{Definitions}\footnote{While the items \textit {g)} and \textit {h)} group definitions by semantic similarity, the item \textit {i)} groups by low occurrence, despite their distinct purposes.}. \textbf{Table \ref{TABLE/severity-levels-definition-libraries}} presents the definitions and descriptions found for the log severity levels on libraries.

\paragraph{\textbf{Debug}} According to the libraries, the \textit{Debug} level describes detailed [L14][L15][L40] and low priority/importance [L04][L05] information, which helps debug activities [L13][L16][L36][L40]. [L15] describes \textit{Debug} as \textit{“too verbose to be included in ‘info’ level,”} suggesting a similarity in the level's purpose. Only one of the libraries uses the word \textit{“problem”} to describe this level [L17].

\paragraph{\textbf{Trace}} The \textit{Trace} level is described with the same characteristics as \textit{Debug} but deepens the low priority: \textit{“very low priority”} [L04], \textit{“very low importance”} [L05].

\paragraph{\textbf{Info}} The \textit{Info} level designates normal behaviors [L16], regular operations [L15], and their messages \textit{“highlight (overall) the progress of the application”} [L05] [L40] \textit{“at coarse-grained level”} [L40]. [L14] advises to keep them to a minimum, as messages at this level will generate data immediately. [L29] highlights that they have value for end users and system administrators. [L17] also uses the word \textit{“problem”} to refer to this level.

\paragraph{\textbf{Warn}} The \textit{Warn} level is described as a \textit{“hazardous situation”} [L04], highlighting a potential problem [L05][L17][L29][L40] that could be to an error [L13]. It is also described as a \textit{“almost errors”} [L14], considering that the application is still running although unexpected [L14][L16]. Operators/end-users/system managers should be likely to be interested in messages at this level [L15][L29].

\paragraph{\textbf{Notice}} The \textit{Notice} level resembles \textit{Info} level in three definitions: it describes normal events [L35] [L36] and highlights the application's progress [L40]. However, according to [L03], it can describe potential failures, likening the \textit{Notice} level to the \textit{Warn} level.

\paragraph{\textbf{Error}} The expressions used to describe the \textit{Error} level are \textit{“major problem”} [L17], \textit{“very serious error”} [L04], \textit{“unexpected conditions”} [L14]. In addition, there are also descriptions that the registered event may or may not interrupt the application's operation [L05] [L16] [L40]. Regardless, even if it does not stop the application as a whole, it can impede the good progress of a particular request [L06]. In the event of logs of this level, an operator must find out as soon as possible [L15]. 

\paragraph{\textbf{Severe, Critical, Alert, Fatal, Emergency}} In the library definitions, five levels appear related to very severe error events: \textit{Severe} [L40], \textit{Critical} [L40], \textit{Alert} [L40], \textit{Fatal} [L13] [L14] [L40], and \textit{Emergency} [L40]. 
\textit{Critical} level definitions lead to believe that a \textit{“disaster”} has occurred [L08].
In addition, \textit{Critical} level requires immediate action [L35] [L36] [L40]. For the \textit{Fatal} level, we find more descriptions for the event: it \textit{“prevent the application from continuing”} [L13], it \textit{“cause premature termination”} [L14], the application \textit{“is going down/to stop”} [L15] [L16], \textit{“lead to abort”} [L40] or \textit{“become unusable”} [L15].

\begin{finding}[label=FINDING:fatal-levels-libraries]
\small
\textit{\textbf{Finding \#\ref{FINDING:fatal-levels-libraries}:} Severe, Critical, Alert, Fatal, and Emergency have similar descriptions regarding severe error events in the libraries.
}
\end{finding}

\paragraph{\textbf{Finest, Verbose, Finer, Fine}}
Descriptions for these levels appear in three  libraries. In [L9], they are described similarly, without precise terms to distinguish their differences, such as the \textit{“most voluminous detailed”}, \textit{“somewhat less detailed”}, and \textit{“lowest volume”} output. For [L12], these levels are useful to debug an application.

\paragraph{\textbf{Basic, Config, Success, Fault}}
These four levels are unique in four distinct libraries. According to [L32], \textit{Basic} is an alias for the \textit{Debug} level. In [L29], \textit{Config} level describes messages \textit{“intended to provide a variety of static configuration information, to assist in debugging problems.”} [L34] is the only library to offer the \textit{Success} severity level, but it does not provide a definition. Numerically, it lies between the \textit{Info} and \textit{Warn} levels. For [L03], \textit{Fault} level \textit{“captures information about faults and bugs in your code”}; the library, as far as we know, does not provide numerical values for the levels. 

\begin{finding}[label=FINDING:convergent-definitions-libraries]
\small
\textit{\textbf{Finding \#\ref{FINDING:convergent-definitions-libraries}:} From the libraries, we found 19 severity levels. Despite the diversity of nomenclature, the concepts of levels are consistent across the various libraries, suggesting a convergence towards concepts of greater granularity.}
\end{finding}

\subsection{Methodology - Practitioners' Point of View}

\bgroup
In Garousi \al 's guidelines \cite{IST/GAROUSI2019/guidelines}, the importance of contextual information in the study suggests the inclusion of grey literature. Therefore, we adopted automated search on \textit{Stack OverFlow}\footnote{ \href{https://stackoverflow.com/}{https://stackoverflow.com/}}, \textit{“a major forum where practitioners post questions and discuss technical issues”} \cite{ICEASE/GAROUSI2016/need-for-multivocal}, as the search strategy for capture the practitioners' point of view. Our search query was: \texttt{log levels}, using the filter \texttt{is:question}. We found \textbf{742 hits} (as of this writing: Jun. 2021). Following, we applied the inclusion (IC) and exclusion (EC) criteria:

\begin{itemize}
    \item \textbf{IC1:} The question/answer explains when to use at least five of the log levels.
    \item \textbf{IC2:} The question/answer must have at least two votes.
    \item \textbf{EC1:} The question/answer is not original (copied from another source such as logging libraries or RFCs).
    \item \textbf{EC2:} The question/answer consists of exemplifying messages characteristic of log levels;
    \item \textbf{EC3:} Questions not approved and closed by Stack Overflow.
\end{itemize}

After  applying  the criteria,  we  obtained \textbf{4 questions} with \textbf{9 relevant answers (Table \ref{TABLE/question-on-stack-overflow})}.

\begin{table}[htbp]
\setlength{\tabcolsep}{0.3em}
\centering
\scriptsize
\caption{Selected questions on Stack OverFlow (QSO)}
\label{TABLE/question-on-stack-overflow}
\begin{tabular}{ c l  l l }
\toprule
\# & Title & URL & Answers  \\
\hline
\lightlightgrayrow {[}QSO1{]}  &  When to use the different log & https://bit.ly/2SQhCE8 & [ASO1][ASO2]\\
\lightlightgrayrow &levels & &[ASO3] \\

{[}QSO2{]}  & Logging levels - Logback  & https://bit.ly/3hNei5d & [ASO4][ASO5]\\
        & rule-of-thumb to assign log levels & & [ASO6][ASO7]\\
        
\lightlightgrayrow {[}QSO3{]}  & Difference between logger.info & https://bit.ly/3hgKKhg & [ASO8] \\
\lightlightgrayrow         & and logger.debug &   &\\
        
{[}QSO4{]}  & How to use log levels in Java &  https://bit.ly/3wa1UBn & [ASO9]\\
\bottomrule
\end{tabular}
\end{table}
\egroup

\subsection{Results - Practitioners' Point of View}

In the selected answers from Stack Overflow, six levels of log severity are described, among which the most discussed are \textit{Debug} (9), \textit{Error} (9), \textit{Warn} (8), and \textit{Info} (8); the other two levels are \textit{Trace} (5) and \textit{Fatal} (3).

\begin{finding}[label=FINDING/main-levels-SO]
\small
\textit{\textbf{Finding \#\ref{FINDING/main-levels-SO}:} The severity levels discussed in the selected responses corroborate the most cited and defined levels in the peer-reviewed literature and logging libraries, respectively.
}\end{finding}

\textbf{Definitions. }
\textbf{Table \ref{TABLE/SO-definitions}} presents the definitions and descriptions found for the log severity levels on libraries.

\bgroup
\setlength{\tabcolsep}{0.07em}
\begin{table*}[bpht]
\centering
\scriptsize
\caption{Severity level definitions from Stack Overflow}
\label{TABLE/SO-definitions}
\begin{tabular}{rcl}
\toprule
Level & Answer & Quote  \\
\hline
\lightlightgrayrow Trace & {[}ASO1{]} & “Only when I would be "tracing" the code and trying to find one part of a function specifically.” \\
 \lightlightgrayrow & {[}ASO2{]} & “Trace is by far the most commonly used severity and should provide context to understand the steps leading up to errors and warnings. (...)” \\
 \lightlightgrayrow & {[}ASO3{]} & “The TRACE messages are intended for developers when they don't need to log state variables.” \\
 \lightlightgrayrow & {[}ASO4{]} & “We don't use this often, (...) extremely detailed and potentially high volume logs that you don't typically want enabled even during normal development. (...)” \\
 \lightlightgrayrow & {[}ASO5{]} & “Trace is something i have never actually used”\\
 
Debug & {[}ASO1{]} & “Information that is diagnostically helpful to people more than just developers (IT, sysadmins, etc.).” \\
 & {[}ASO2{]} & “We consider Debug $<$ Trace. (...) we discourage use of Debug messages (...) this makes log files almost useless (...)” \\
 & {[}ASO3{]} & “The DEBUG messages are intended for developers when they need to log state variables.” \\
 & {[}ASO4{]} & “(...) any message that is helpful in tracking the flow through the system and isolating issues, especially during the development and QA phases. (...)” \\
 & {[}ASO5{]} & “Debug means that something normal and insignificant happened; (...)” \\
 & {[}ASO6{]} & “Shouldn't be used at all (and certainly not in production) (...)” \\
 & {[}ASO7{]} & “variable contents relevant to be watched permanently” \\
 & {[}ASO8{]} & “If you want to print the value of a variable at any given point, you might call Logger.debug” \\
 & {[}ASO9{]} & “As the name says, debug messages that we only rarely turn on. (...)” \\
 
\lightlightgrayrow Info & {[}ASO1{]} & “Generally useful information to log (service start/stop, configuration assumptions, etc). (...) I want to always have available but usually don't care about under\\
 \lightlightgrayrow & & normal circumstances. This is my out-of-the-box config level.” \\
 \lightlightgrayrow & {[}ASO2{]} & “This is important information that should be logged under normal conditions such as successful initialization, services starting and stopping or successful \\
 \lightlightgrayrow & & completion of significant transactions. (...)” \\
 \lightlightgrayrow & {[}ASO3{]} & “The INFO messages are intended for system operators and describe expected states” \\
 \lightlightgrayrow & {[}ASO4{]} & “Things we want to see at high volume in case we need to forensically analyze an issue. System lifecycle events (system start, stop) go here. (...) Typical \\ 
 \lightlightgrayrow & & business exceptions can go here (...)” \\
 \lightlightgrayrow & {[}ASO5{]} & “Info means that something normal but significant happened; the system started, the system stopped, (...)” \\
 \lightlightgrayrow & {[}ASO6{]} & “Anything else that we want to get to an operator.(...) log message per significant operation (...).” \\
 \lightlightgrayrow & {[}ASO7{]} & “used in functions/methods first line, to show a procedure that has been called or a step gone ok, (...)” \\
 \lightlightgrayrow & {[}ASO9{]} & “Anything that we want to know when looking at the log files, e.g. when a scheduled job started/ended (...)” \\
 
Warn & {[}ASO1{]} & “Anything that can potentially cause application oddities, but for which I am automatically recovering. (...)” \\
 & {[}ASO2{]} & “This MIGHT be problem, or might not. (...) Viewing a log filtered to show only warnings and errors may give quick insight into early hints at the root cause\\
 & & of a subsequent error. Warnings should be used sparingly so that they don't become meaningless. (...)” \\
 & {[}ASO3{]} & “The WARN messages are intended for system operators when the process can continue in an unwanted state” \\
 & {[}ASO4{]} & “An unexpected technical or business event happened, customers may be affected, but probably no immediate human intervention is required. (...) Basically any\\
 & & issue that needs to be tracked but may not require immediate intervention.” \\
 & {[}ASO5{]} & “Warn means that something unexpected happened, but that execution can continue, perhaps in a degraded mode;(...) Something is not right, but it hasn't gone\\
 & & properly wrong yet - warnings are often a sign that there will be an error very soon.” \\
 & {[}ASO6{]} & “This component has had a failure believed to be caused by a dependent component (...). Get the maintainers of THAT component out of bed.” \\
 & {[}ASO7{]} & “not-breaking issues, but stuff to pay attention for. Like a requested page not found” \\
 & {[}ASO9{]} & “Any message that might warn us of potential problems, (...)” \\
 
\lightlightgrayrow Error & {[}ASO1{]} & “Any error which is fatal to the operation, but not the service or application (...) These errors will force user (administrator, or direct user) intervention. (...)” \\
 \lightlightgrayrow & {[}ASO2{]} & “Definitely a problem that should be investigated. SysAdmin should be notified automatically, but doesn't need to be dragged out of bed. (...)” \\
 \lightlightgrayrow & {[}ASO3{]} & “The ERROR messages are intended for system operators when, despite the process cannot continue in an unwanted state, the application can continue.” \\
 \lightlightgrayrow & {[}ASO4{]} & “The system is in distress, customers are probably being affected (or will soon be) and the fix probably requires human intervention. The "2AM rule" applies \\
 \lightlightgrayrow & & here-if you're on call, do you want to be woken up at 2AM if this condition happens? If yes, then log it as ‘error’” \\
 \lightlightgrayrow & {[}ASO5{]} & “Error means that the execution of some task could not be completed; (...) Something has definitively gone wrong.” \\
 \lightlightgrayrow & {[}ASO6{]} & “This component has had a failure and the cause is believed to be internal (...). Get me (maintainer of this component) out of bed.” \\
 \lightlightgrayrow & {[}ASO7{]} & “critical logical errors on application, like a database connection timeout. Things that call for a bug-fix in near future” \\
 \lightlightgrayrow & {[}ASO8{]} & “When responding to an Exception, you might call Logger.error” \\
 \lightlightgrayrow & {[}ASO9{]} & “Any error/exception that is or might be critical. Our Logger automatically sends an email for each such message on our servers” \\
 
Fatal & {[}ASO1{]} & “Any error that is forcing a shutdown of the service or application to prevent data loss (or further data loss).” \\
 & {[}ASO2{]} & “Overall application or system failure that should be investigated immediately.(...) wake up the SysAdmin. (...) this severity should be used very infrequently(...)” \\
 & {[}ASO3{]} & “The FATAL messages are intended for system operators when the application cannot continue in an unwanted state.” \\
 \bottomrule
\end{tabular}

\end{table*}
\egroup
\paragraph{\textbf{Debug and Trace}}
Compared to the first two sources in our mapping, the similarity observed between the severity levels is not as striking in the responses selected from Stack Overflow. It would be best to prefer \textit{Debug} over \textit{Trace} for part of the answers [ASO1][ASO4][ASO5], and the opposite for another part [ASO2][ASO6][ASO9]. For [ASO3], both levels are intended for developers, \textit{Debug} being the one that records variable values. For [ASO1], \textit{Trace} level is used to find a specific piece of code, while Debug level is classified as \textit{“helpful to people more than just developers.”}

\paragraph{\textbf{Info}}
Among the levels discussed on the selected answers, the \textit{Info} level has the most significant convergence in the definitions presented. All answers describe it as a record of operations that start and/or end, describing \textit{“normal but significant situations”} of a software system {[}ASO1{]}{[}ASO2{]}{[}ASO5{]}{[}ASO6{]}, \ie \textit{“expected situations”} {[}ASO3{]}. {[}ASO4{]} points out that this level also describes typical business exceptions, and according to {[}ASO6{]}, operators are the audience of \textit{Info} messages. 

\paragraph{\textbf{Warn}}
As in the libraries, the selected answers describe the \textit{Warn} level messages as potential problems/unexpected events {[}ASO1{]}{[}ASO2{]}{[}ASO4{]}{[}ASO5{]}{[}ASO9{]} that can cause complications for the  system {[}ASO1{]}{[}ASO5{]}, and therefore they need to be observed. Despite these events, the system remains running {[}ASO1{]}{[}ASO3{]}{[}ASO5{]}, without the need for immediate human intervention {[}ASO4{]}. For {[}ASO3{]}, operators are the public interested in this level of severity.

\paragraph{\textbf{Error, Fatal}}
There is also divergence at this level. For four  of the responses {[}ASO1{]}{[}ASO2{]}{[}ASO3{]}{[}ASO6{]}, the \textit{Error} level indicates a failure that did not stop the system execution but should be investigated by the system operators {[}ASO2{]}{[}ASO3{]}. However, for another two responses, the degree of severity is more critical, and the \textit{“interested person”} should be \textit{“get out of bed”} {[}ASO4{]}{[}ASO5{]}. This severity degree is the same that is attributed to the \textit{Fatal} level by {[}ASO1{]}{[}ASO2{]}{[}ASO3{]}: errors occur that force the application to \textit{“shut down”} and require immediate action.

\section{Log severity level synthesis}
\label{SECTION/combined-results}
Analyzing our three sources, we observe redundancy in the numerical values of the levels, the semantic similarity of their definitions, and the low occurrence of some levels in the libraries. To reduce the similar or redundant levels, we abstracted the 19 severity levels to six levels, which constitute the logging's state of the practice. Following, we explain the steps of our synthesis, as shown in \textbf{Fig. \ref{IMAGE/levels}}.

\subsection{Abstracting of Log Severity Levels}
\label{SUBSECTION/abstraction-synthesis}

\begin{figure*}[htbp]
\centerline{\includegraphics[width=0.8\linewidth]{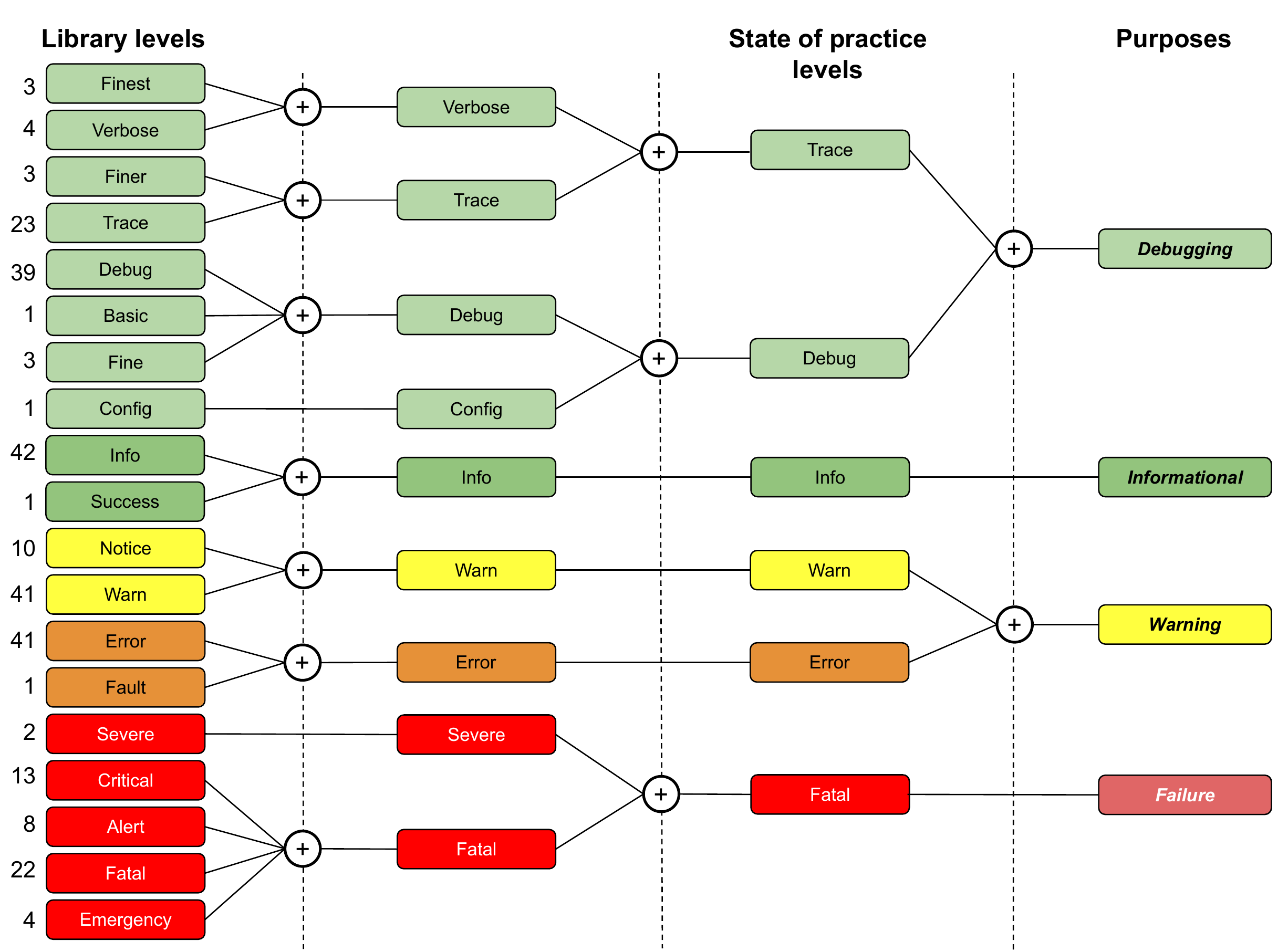}}
\caption{Abstraction of log severity levels: Considering the semantic similarity, the numerical redundancies of value, and the occurrence in the libraries of the severity levels, we converged 19 levels to six levels. The six levels can be associated with three  log severity purposes (The numbers to the left of the level indicate the occurrence of each level in the selected libraries).}
\label{IMAGE/levels}
\end{figure*}

\textit{\textbf{Finest, Verbose, Finer, Trace, Debug, Basic, Fine, Config.}}
\textit{Finest} and \textit{Verbose} levels are present in six libraries. [L29] and [L32] provide the \textit{Finest} level; [L10], [L12], and [L19] provide the \textit{Verbose} level, and [L40] provide both levels. Besides the semantic similarity between their definitions/descriptions in six libraries, in [L40] they (\textit{Finest, Verbose}) have the same numerical value. These facts suggest they can be merged on the same level. In \textbf{Fig. \ref{IMAGE/levels}}, we chose to merge for the severity with the highest occurrence. We performed the same process for \textit{Finer} and \textit{Trace}, as well as \textit{Debug} and \textit{Fine} levels. The \textit{Basic} level, present only in [L32], is equated with the \textit{Debug} severity level in its documentation and merges to \textit{Debug}. The resulting \textit{Verbose} and \textit{Trace} still have definitions with substantial semantic similarity, so we abstracted the level from \textit{Verbose} to \textit{Trace}. The low occurrence \textit{Config} level, which provides information for debugging [L29], was merged with the \textit{Debug} level.

\textit{\textbf{Info, Success.}}
We observed that the \textit{Success} level has a numerical value between \textit{Info} and \textit{Warn} levels in [L34]. Besides, the \textit{Success} level has low occurrence and does not have a definition. We merged into \textit{Info} as it goes against \textit{Warn}’s purpose.

\textit{\textbf{Notice, Warn.}}
In libraries, the \textit{Notice} level has low occurrence. Also, it resembles the \textit{Info} level at the same time that it resembles the \textit{Warn} level. Considering the combination of its definition found in the literature and the level nomenclature, we merged the \textit{Notice} level to the \textit{Warn} level.

\textit{\textbf{Error, Fault.}}
The \textit{Fault} level is another level of low occurrence. For [L03], \textit{Fault} level describes bugs in running software systems, and furthermore, it ranks before the \textit{Critical} severity level. Consequently, we merged \textit{Fault} to \textit{Error}.

\textit{\textbf{Fatal, Alert, Critical, Emergency, Severe.}} 
\textit{Fatal}, \textit{Critical}, \textit{Alert}, and \textit{Emergency} levels are the most severe levels
out of 81\% of selected libraries, and their definitions have a prominent semantic similarity. In our abstraction, the \textit{Fatal} level is the level that describes the failures situation, and it has the most significant among the levels analyzed (\textit{Fatal}, \textit{Critical}, \textit{Alert}, \textit{Severe} and \textit{Emergency}). The \textit{Fatal} and \textit{Critical} levels have the same value in [L17].  Finally, the \textit{Severe} level has low occurrence, but its definition has semantic similarity with the \textit{Fatal} level. These facts leaded us to merge \textit{Fatal}, \textit{Critical}, \textit{Alert}, \textit{Severe} and \textit{Emergency} into the \textit{Fatal} level.

\subsection{Synthesized Definitions}
\label{SECTION/synthesized-definitions}

We performed the synthesis process, resulting on six state-of-the-practice  log severity levels: \textit{Trace, Debug, Info, Warn, Error, Fatal}. Thus, considering the results obtained from the three  sources, we synthesized combined definitions for the six abstracted levels as follows.

\begin{finding}
\textit{\textbf{Debug severity level} describes variable states and details about interesting events and decision points in the execution flow of a software system, which helps developers to investigate internal system events.
}\end{finding}

\begin{finding}
\textit{\textbf{Trace severity level} broadly tracks variable states and events in a software system.}
\end{finding}

\begin{finding}
\textit{\textbf{Info severity level} describes normal events, which inform the expected progress and state of a software system.} 
\end{finding}

\begin{finding}
\textit{\textbf{Warn severity level} describes potentially dangerous situations caused by unexpected events and states. For this reason, they must be observed, even if they do not interrupt the execution of a software system.
}\end{finding}

\begin{finding}
\textit{\textbf{Error severity level} describes the occurrence of unexpected behavior of a software system. For this reason, they must be investigated, even if they do not interrupt the execution of a software system.
}\end{finding}

\begin{finding}
\textit{\textbf{Fatal severity level} describes critical events that bring a software system to failure.
}\end{finding}

\subsection{Purpose of Log Severity Levels}

We observed a convergence of log severity levels across the three sources after the processes of abstraction and synthesis, and we noticed \textbf{four main purposes for log severity levels}: 

\paragraph{\textit{Debugging Purpose}} it describes levels used to log variable states and events internal to the behavior of a software system. It groups \textit{Debug} and \textit{Trace}, while \textit{Trace} extrapolates \textit{Debug}'s characteristics of describing variables and events.

\paragraph{\textit{Informational Purpose}}
it describes levels used to record the expected behavior of a software system.

\paragraph{\textit{Warning Purpose}} it describes levels used to warn unexpected behavior of a software system. It groups \textit{Error} and \textit{Warn} levels because both indicate problems (or potential problems) that should be investigated, but do not interrupt the system's execution.

\paragraph{\textit{Failure Purpose}} it describes levels used to record failures of a software system.

\section{Discussion}
\label{SECTION/discussion}

\subsection{There is an excessive variety of severity levels among logging libraries.}
Two-thirds (68\% -- 13/19) of log severity levels can be considered redundant specializations of the recurrent severity levels (32\% -- 6/19) in peer-reviewed literature, logging libraries, and practitioners' point of view. 
The excessive specialization of levels makes its choice difficult, impacting the amount of log generated and, consequently, the reliability of monitoring systems.
We recommend keeping a consistent nomenclature for the levels, and \textbf{we recommend using six severity levels: \textit{Trace}, \textit{Debug}, \textit{Info}, \textit{Warn}, \textit{Error}, and \textit{Fatal}}.
We also recommend creating a policy to use severity levels, with practical examples to guide the choice of severity levels effectively.

\subsection{There is a lack of precision in the definitions of log severity levels.}
We observed a lack of precision in the definitions of log severity levels. For example, there are severity levels in library definitions without distinguishing specific purposes or only characterized by adjectives and superlatives. This lack of precision can cause a misunderstanding of severity levels and hinder logging practices. Thus, \textbf{we suggest that logging library creators provide precise and unambiguous definitions for considering the log level purposes.}

\subsection{The values of log severity levels help to understand the distinction between severity levels.}
The vast majority of logging libraries use a numeric value associated with severity levels. In the absence of precise definitions, these values clarify the library creators' proposal regarding the degree of severity of each level. \textbf{We suggest that logging library creators continue to provide these values. We also suggest that these values are ascending, considering the order between the six severity levels of the state of the practice,} from the least severe to the most severe: \textit{Trace} $<$ \textit{Debug} $<$ \textit{Info} $<$ \textit{Warn} $<$ \textit{Error} $<$ \textit{Fatal}.

\subsection{Explore the logging library features.}
Using a limited set of levels can lead to difficulties when there is a need to identify specific log data. \textbf{We recommend that developers explore the library features like the \textit{“Marker interface”}, } which can add semantics to the log levels used in practice. \textbf{We also recommend studying the logging library settings to use configuration practices instead specialized severity levels.}.

\subsection{Research  on  log  severity  levels  has  grown  in  recent years.}
The logging community recognizes that choosing the level of log severity can be challenging and can impact systems in development and production.
Therefore, in recent years, log severity levels have been more generally investigated. However, despite this growing interest, the literature had not established what are log severity levels indeed. Thus, our study is a first step towards establishing a systematic mapping of severity levels, definitions and purposes.

\subsection{Threats to validity}
\label{SECTION/threats-to-validity}

\paragraph{Validation of definitions and purposes of severity levels} 

In this study, we do not validate definitions and purposes against actual logging entries. Therefore, empirical studies need to be done to observe the adherence of these definitions besides logging practices in software systems. Moreover, controlled experiments should be performed to assess the effectiveness of these definitions.

\paragraph{Logging library mapping} The fact that our work does not cover an exhaustive set of libraries is a factor that can reduce the validity of the results. Nevertheless, we aimed to obtain a representative set of them, applying consistent inclusion and exclusion criteria, selecting only libraries to increase the validity of our results to mitigate this threat.

\section{Conclusion}
\label{SECTION/conclusion}

The choice of log severity level can be challenging and cause problems in producing reliable logging data. In this study, we present a state-of-the-art and state-of-the-practice mapping of log severity levels. We extracted data from three sources: peer-reviewed literature, logging libraries, and practitioners' point of view. Our study systematically mapped the selected sources, empirically analyzed the definitions, descriptions, and documentation of log severity levels.
To summarize, we analyzed 19 severity levels from 27 studies and 40 logging libraries. Our results showed that there is redundancy and semantic similarity between the levels. Moreover, \textbf{they converge the severity levels for a total of six levels:\textit{ Trace, Debug, Info, Warn, Error and Fatal}}. Besides, there is consistency in ordering between the different levels in different libraries and that the levels are permeated with specific purposes.  

Our main contributions are:
(i) mapping of the peer-reviewed literature of studies dealing with logging severity levels;
(ii) mapping of the severity levels in the logging libraries;
(iii) a set of synthesized definitions for log severity levels, and four general purposes for severity levels.
The results of our study (mapping, definitions, and purposes) provide evidence to create guidelines for choosing log severity levels that increase the data reliability. Furthermore, logging library creators can use our results to adopt severity levels accordingly. Finally, we present recommendations about log severity levels.

In future work, we plan to expand this systematic multivocal mapping. We will add more logging libraries and grey literature sources,  other Q\&A websites, discussions and technical blogs that discuss the log severity level.
Furthermore, we aim to leverage the results of this mapping, identifying a conceptual framework that supports developers and system operators' logging practices in addition to metrics and approaches. Finally, we plan to organize a catalogue of log entry patterns, presenting metadata for each severity level as intent and practical examples.

\balance
\bibliographystyle{IEEEtran}
\bibliography{references}

\vspace{12pt}

\end{document}